\documentclass[review]{elsarticle}

\usepackage[normalem]{ulem} 
\usepackage{etoolbox,lineno}

\usepackage{amsmath}
\usepackage[font=footnotesize,labelfont=bf]{subcaption}

\begin{document}

\begin{frontmatter}
\title{Blockchain-Based Process Control and Monitoring Architecture for Vertical Integration of Industry 4.0}
 

\author[mymainaddress,mysecondaryaddress]{Charles T. B. Garrocho\corref{mycorrespondingauthor}}
\cortext[mycorrespondingauthor]{Corresponding author}
\ead{charles.garrocho@ifmg.edu.br}

\author[mysecondaryaddress]{Célio M. S. Ferreira}
\author[mysecondaryaddress]{Carlos F. M. C. Cavalcanti}
\author[mysecondaryaddress]{Ricardo A. R. Oliveira}

\address[mymainaddress]{Minas Gerais Federal Institute of Education, Science and Technology}
\address[mysecondaryaddress]{Computer Science Department, Federal University of Ouro Preto}

\begin{abstract}
Industrial Internet of Things is a new milestone that will require new industry paradigms and investments. In this context, cyber-physical systems are considered the bridge to the fourth revolution. Centralized approaches and observance of real-time constraints are two important challenges that must be overcome for the advancement of Industry 4.0. To solve these problems, a blockchain-based vertical integration architecture of the process automation systems is proposed in which it performs the control and monitoring of industrial processes. Proof of concept experiments reveal the feasibility and performance of the proposal.
\end{abstract}


\end{frontmatter}

\section{Introduction}

Industry 4.0 refers to the fourth industrial revolution that transforms industrial manufacturing systems into cyber production systems, introducing emerging paradigms of information and communication, such as the Internet of Things (IoT) \cite{serpanos2018industrial}. By 2025, Industrial IoT (IIoT) investments in the world should reach US\$ 949.42 billion \cite{Reportbuyer2019}. With the introduction of Industrial IIoT in the factory, a 30\% increase in productivity is expected, generating a very optimistic investment forecast of US\$ 13 trillion by 2030 \cite{accenture2015inc}.

\newpage

Cyber-physical systems are the enablers of the new industrial age that integrate factory processes with business decisions and provides the link between customers and suppliers \cite{colombo2017industrial}. Besides, cyber-physical systems enable vertical integration of Process Automation Systems (PAS) by introducing field device communication to decision-making levels. In this new context, systems are designed for IIoT devices that have unique features such as low processing power and storage capacity, low bandwidth for data transmission and collection, and limited autonomy \cite{sisinni2018industrial, garrocho2020d2d}.

Despite the many advantages that the introduction of cyber-physical systems can offer, some problems must be overcome: Machine-to-machine (M2M) communication based on Publish-Subscribe paradigm uses a communication model through an intermediate node that becomes a point of failure \cite{wagner2017industry}; M2M communication latency can affect the time requirements of real-time systems, compromising deadlines \cite{garrocho2020real}; industrial plant updates with new equipment that are expensive and unrealistic for small and medium industries and will replace those that already work perfectly \cite{herterich2015impact}.

To work around these issues, a blockchain-based PAS vertical integration architecture is presented. This proposal allows a decentralized M2M network to be used without the need for a trusted broker, ensuring that all processes run reliably and without change; minor changes to the existing Programming Logic Control (PLC) at the plant to avoid the purchase of new equipment and major changes at the plant; the operation of the proposed approach allows data collected from sensors and actuator actions to be recorded on the blockchain network, operating safely, traceable and without compromising system requirements in real-time.

The remainder of this paper is organized as follows: Section \ref{sec:ant} introduces PAS, blockchain, smart contract, and discusses related work. Section \ref{sec:pro} presents the architecture and operation of the proposal. Section \ref{sec:ava} presents the proof of concept, the scenario, and the evaluation metrics of the experiments. Section \ref{sec:ana} analyzes and discusses the results of the experiments. Finally, Section \ref{sec:con} presents the final considerations.

\section{Background}
\label{sec:ant}

At this beginning of the fourth industrial revolution, the role of industrial networks is becoming increasingly crucial as they are expected to meet the new and more demanding requirements in any new operating context \cite{wollschlaeger2017future}. IIoT networks are usually used to monitor conditions, manufacturing processes and predictive maintenance. Such networks have typical configurations, traffic, and performance requirements that make them distinct and different from traditional communication systems generally adopted by general-purpose applications. The most critical requirements are time, reliability and flexibility \cite{felser2005real}.


\subsection{PAS Hierarchy and Synchronicity Requirements}

\begin{figure}[!h]
\centering
\includegraphics[width=0.8\linewidth]{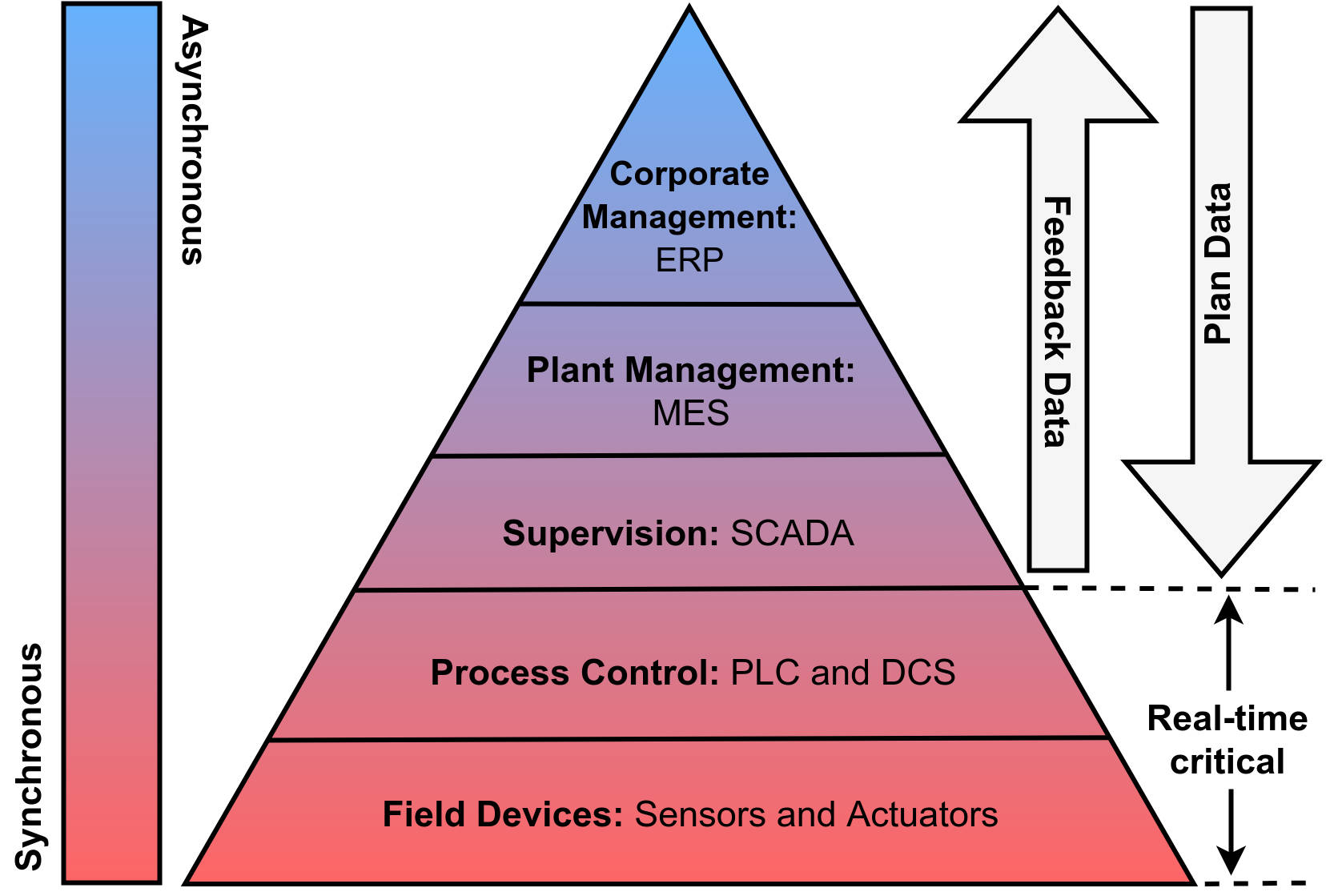}
\caption{PAS hierarchy levels and synchronicity requirements.}
\label{fig:hierarquia}
\end{figure}

Unlike many network protocols and information systems already widely adopted in homes and commerce, PAS uses protocols and systems designed specifically to be used in industry environments. As showed in Figure \ref{fig:hierarquia}, PAS are typically based on a five-level hierarchy \cite{mehta2014industrial}. These systems are generally adopted by continuous industrial processes such as oil and gas distribution, power generation and management, chemical processing and treatment of glass and minerals.

The field devices level are composed by sensors and actuators controlled by the control level. Control level consists of devices that provide an interface for Internet Protocol (IP)-based network communication at the supervision level such as Programmable Logic Controller (PLC) and Distributed Control System (DCS) \cite{vitturi2019industrial}. At the supervision level, processes are monitored and executed by operators through systems like the Supervision Control and Data Acquisition (SCADA). Finally, in the last layers, there is corporate and plant management and sending process-related data to the cloud, trough systems like the Manufacturing Execution System (MES) and Enterprise Resource Planning (ERP).

PAS comprises many nodes, logically positioned at various hierarchical levels and distributed over large geographical areas \cite{sharma2016overview}. Many Human-Machine Interface (HMI) servers and computers are used for interaction between man and the control level. As presented in Figure \ref{fig:hierarquia}, at the bottom of the pyramid, the processes are mostly synchronous and real-time critical. The processes presented at the top of the pyramid are goal-oriented and mainly asynchronous.

\subsubsection{Vertical Integration of PAS in Industry 4.0}

Vertical integration in Industry 4.0 targets to unite all logical layers of the organization, starting from field layers  (i.e. the production area) to more abstract and higher layers as quality assurance, product management, sales, marketing, etc \cite{perez2019vertical}. In these layers, data flows freely and transparently up and down in order to help produce strategic and tactical data-driven decisions. The Industry 4.0 approach integrates enterprise gains vertically in a crucial competitive advantage by being able to respond appropriately and quickly to any  market changed signals and new opportunities.

Data sharing is a requirement for vertical integration, but it is not an easy task \cite{hvi2019i40}. It starts at the production level, where various equipment from various manufacturers can be found, with varying levels of automation, and equipped with a wide range of sensors and using different communication protocols. In other words, they usually “don't speak the same language” and it is necessary to establish a meta-network that addresses these communication disparities.

According to Nilsson et al. \cite{nilsson1998stochastic}, PAS have evolved into a technology stage in which they are distributed and controlled using the network. Therefore, there is a delay related to network flow of data that results in a challenge in the PAS vertical integration process \cite{garrocho2020performance}. This challenge becomes even greater and quite dangerous when processes are being controlled by real-time systems where applications are time-critical and have deadline restrictions. In modern industrial approaches, this poses challenges in communicating with the concepts of industrial cyber-physical systems \cite{jeschke2017industrial}.

Cyber-physical systems are enablers of the new industrial age integrating processes from the operational and field level to business decisions level. Jeschke et al. \cite{jeschke2017industrial} points out that the concept of industry 4.0 is a proposition of a fourth industrial revolution based on Internet connections, allowing the integration and cooperation of manufacturing machines. However, this scenario presents a challenge because industrial processes depends on synchronous elements and real-time processing. Real-time processing is a system-level requirement in new industrial devices connected to the Internet, the IIoT devices \cite{pinto2017iioteed}.

\subsection{Blockchain and Smart Contracts}

IIoT devices are expected to have a long service life for being used in an industrial infrastructure without creating vulnerabilities. Machines equipped with IIoT devices that will be responsible for exchanging goods or services must use secure M2M transactions to provide risk-free and fault-tolerant operation \cite{hill2019securing}. Some blockchain functionality can ensure reliable and decentralized M2M communications, in which network nodes do not need a trusted intermediary to exchange messages with each other or with a central authority \cite{garrocho2019industry, fernandez2019review}.

Blockchain is a decentralized P2P network that has excellent fault tolerance. Blockchain transactions cannot be deleted or changed. Blockchain is highly scalable, and all transactions are encrypted, making them secure, auditable, and transparent. At the heart of this technology, there are the consensus algorithms, which are protocols for obtaining data value agreements between nodes distributed across the network \cite{banerjee2018blockchain}. Consensus algorithms are designed to achieve reliability in a network that involves multiple untrusted nodes. Proof of Work (PoW), Proof of Stake (PoS), and Proof of Authority (PoA) are currently the most used algorithms.

Blockchain can be permissionless or permissioned. In permissionless approach, transactions are validated by public nodes. In permissioned blockchain, transactions are validated by a select group of nodes approved by the blockchain’s owner  \cite{wust2018you}. Permissioned systems tend to be more scalable and faster, but are more centralized. Permissionless systems are open for all nodes to participate and thus provide a more decentralized approach where the trade-off is speed and scalability. Bitcoin, the best known digital currency uses a blockchain-based permissionless distributed ledger that maintains the transaction history of the bitcoin \cite{zheng2018blockchain}. After the bitcoin, new blockchain-based applications emerged.

\begin{figure*}[h]
\centering
\includegraphics[width=0.98\textwidth]{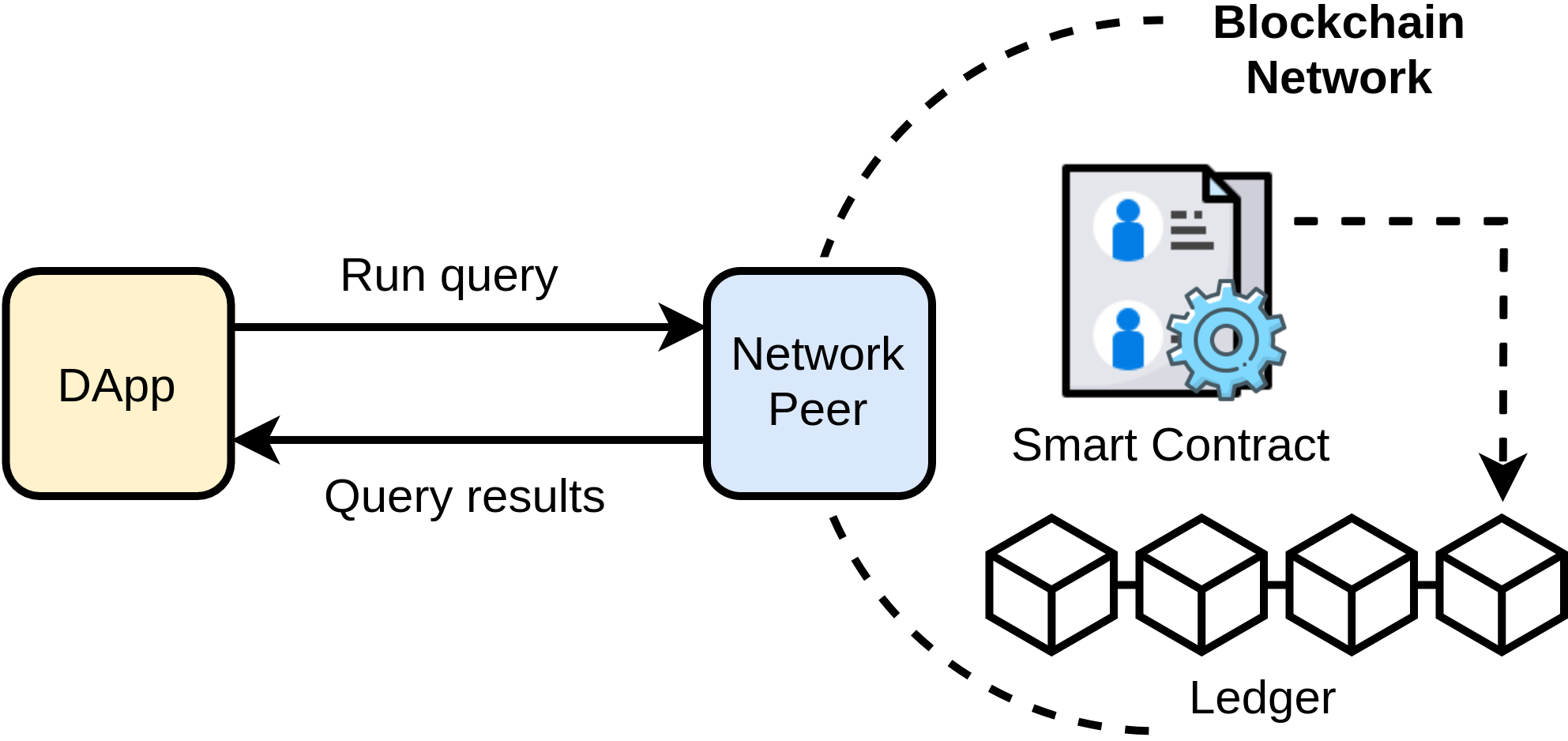}
\caption{Blockchain-based smart contracts operation overview.}
\label{fig:func_sc}
\end{figure*}

The blockchain-based smart contract is a new approach intended to digitally facilitate, verify, or enforce the negotiation or performance of a contract \cite{miller2018smart}. Smart contracts can transact between different nodes without the intermediation of a third-party entity or agent. Ethereum, Hyperledger (Fabric, Sawtooth), and Corda are popular smart contract platforms that are contributing significantly to the generation of Decentralized Applications (DApp) \cite{voulgaris2019blockchain}. As illustrated in Figure \ref{fig:func_sc}, DApps queries the blockchain network through a network peer that executes the smart contract for ledger access.

\subsection{Related Work}

An industrial plant implements networks to connect production systems vertically. A vertical connection is a type of link between two entities that participate in the value chain of a product. Therefore, when this connectivity is automated, information can be automatically collected and sent from the various systems deployed in a factory. Traditionally, this type of integration occurs through manufacturing systems, product lifecycle management, and resource planning. However, Industry 4.0 requires higher levels of integration. Blockchain is the key to improving integration, capturing, processing and returning reliable data through which various plant entities can interact.

To identify works that perform vertical integration of the PAS hierarchy and blockchain-based solutions in this context, research was conducted. Following the Kitchenham \cite{kitchenham2004procedures} protocol, we conducted searches between June and August 2020 on computer databases: ACM Digital Library, Google Scholar, IEEE Xplore Digital Library, and ScienceDirect (Elsevier). To search, we used the string \textsc{(process AND automation AND system AND vertical AND integration AND (factory OR industry)) OR (blockchain AND smart contract AND (factory OR industry))}. Regarding the selection criteria of the articles found, only those that provided a full text in English and published less than four years ago were used.

As a result of this systematic review, we filtered and selected 19 articles from the 320 found. These 19 articles are listed and sorted in Table \ref{tabela:works}. This table also shows the characteristics of each work and groups them according to five columns that we separate as follows: the level of Industrial Plant Changes related to the necessary equipment and software, which may be high, medium or low; Communication Architecture used between the devices, which can be centralized (single-node based) or decentralized (multi-node based); Proposal with or without blockchain-based operation, in which smart contracts are used; Vertical Integration of PAS hierarchy levels that can be full, partial or none; Deadlines Can be Influenced by real-time systems by controlling and executing the proposed system architecture.

\begin{table}[]
\centering
\resizebox{\textwidth}{!}{%
\begin{tabular}{cccccc}
\hline
\textbf{Work} & \rotatebox[origin=c]{90}{\textbf{\begin{tabular}[c]{@{}c@{}}Industrial\\Plant Change\end{tabular}}} &
\rotatebox[origin=c]{90}{\textbf{\begin{tabular}[c]{@{}c@{}}Communication\\Architecture\end{tabular}}} & \rotatebox[origin=c]{90}{\textbf{\begin{tabular}[c]{@{}c@{}}Blockchain\\Based\end{tabular}}} & \rotatebox[origin=c]{90}{\textbf{\begin{tabular}[c]{@{}c@{}}Vertical\\Integration\end{tabular}}} & \rotatebox[origin=c]{90}{\textbf{\begin{tabular}[c]{@{}c@{}}Deadlines Can\\be Influenced\end{tabular}}}\\\hline
\cite{alexakos2017exposing, alexakos2018iot, shirazi2019cloud} & Medium & Centralized & No & Yes & Yes \\ 
\cite{garcia2016opc, calderon2018integration, llamuca2019integrating, liu2019cyber} & High & Centralized & No & Yes & No \\
\cite{kapitonov2018blockchain, petroni2018big, liang2018reliable, schulz2018multichain, vatankhah2019blockchain} & High & Decentralized & Yes & No & No \\
\cite{leang2018real, maw2019ics} & Low & Decentralized & Yes & Partially & No \\ 
\cite{smirnov2018robot, gallo2018decymo, isaja2018distributed, lee2019blockchain, petroni2019blockchain} & High & Decentralized & Yes & Yes & Yes \\
\begin{tabular}[c]{@{}c@{}}Proposed\\Architecture\end{tabular} & Medium & Decentralized & Yes & Yes & No \\ \hline
\end{tabular}%
}
\caption{Classification and grouping of related works.}
\label{tabela:works}
\end{table}

Industry 4.0 aims for a highly flexible and digitized industrial production model that is smarter and more reliable than today's possibilities. This requires vertical integration of different operations in one manufacturing to promote a reconfigurable intelligent factory. Using raw data as an asset from which value can be created to support business and manufacturing decisions has motivated many scientists to explore the challenges of how to exploit that value. In this context, works like \cite{pal2017model} introduce optimization models to schedule maintenance operations using formal methods. However, this approach does not indicate how raw data can be collected and transmitted to the top of the PAS hierarchy.

To integrate IoT into the manufacturing process, some work \cite{alexakos2017exposing, alexakos2018iot, shirazi2019cloud} presents approaches that combine field device networking and high-level multi-agent systems that contribute to vertical integration. However, the evaluation of such proposals was by simulation only, and it is not considered the impact of IP-based networks that can occur on real-time system deadlines. Other works \cite{garcia2016opc, calderon2018integration, llamuca2019integrating, liu2019cyber}, based on the Open Platform Communications Unified Architecture (OPC-UA) protocol, presents vertical integration architectures whose temporal requirements of real-time systems can be met. However, the need for expensive equipment (unrealistic for small and medium industries) and a server to intermediate communication between the components of these architectures makes such architectures vulnerable in the event of a centralized server failure.

Aiming at making communication and decision making decentralized, a lot of works is applying blockchain technologies in the industry. Blockchain-based smart contracts are being applied across supply chains to improve decision making in control and management processes \cite{kapitonov2018blockchain, petroni2018big}, trusted data generation and privacy \cite{liang2018reliable}, reliable communication, and between end-user and service provider \cite{schulz2018multichain, vatankhah2019blockchain}. However, such approaches aim at horizontal communication between companies and/or customers. Aiming for vertical integration of the PAS hierarchy, the works \cite{leang2018real, maw2019ics} feature blockchain-based architectures for cyber-physical systems.

The approaches proposed by the works \cite{leang2018real, maw2019ics} are intended only to monitor data from field devices controlled by the PLC and to record this information on the blockchain to generate an unchanging history. The works \cite{smirnov2018robot, gallo2018decymo, isaja2018distributed, lee2019blockchain, petroni2019blockchain} have similar approaches in which the objective is to use blockchain-related technologies to control processes involving devices and businesses. However, the process control of these architectures is implemented through smart contracts, where the execution time is variable (either by network latency or by committing transactions on the blockchain network), making processes unsafe and prone to failures to meet deadlines in real-time systems.

As shown in Table \ref{tabela:works}, the architecture proposed in this paper differs from other approaches in that it has a decentralized, blockchain-based communication architecture in which the operation of vertical integration of PAS hierarchy levels does not influence the deadlines of real-time systems. To achieve these goals, the architecture is designed to take advantage of all the operation and communication of familiar and widely used equipment (such as the PLC) by incorporating the blockchain into the task orders (such as compilation, execution, and monitoring) performed by the operator at the level from HMI. Besides, the architecture reuses existing equipment and allows few changes to equipment in the industrial plant, requiring less investment by the industry.

\newpage

\section{Blockchain-based Control and Monitoring Architecture Design}
\label{sec:pro}

\begin{figure*}[h]
\centering
\includegraphics[width=\textwidth]{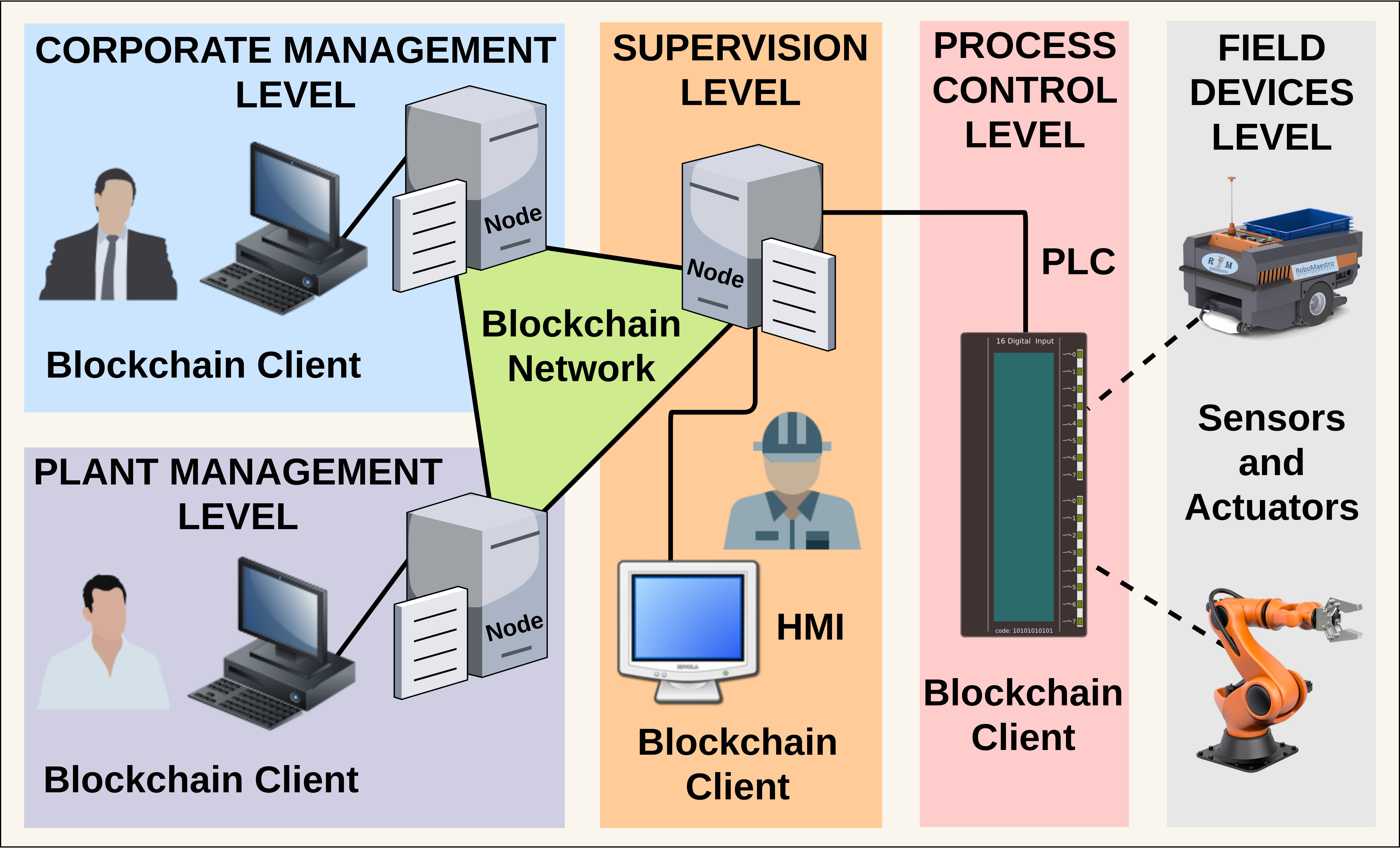}
\caption{Blockchain network positioning in the PAS hierarchy.}
\label{fig:arquitetura}
\end{figure*}

Industry 4.0 technologies can benefit from the use of smart contracts, but their application also presents challenges in many ways. Deployment a blockchain can help cloud-based solutions provide redundancy for storage needs, while at the same time this local blockchain deployment is currently challenging to replicate on IIoT nodes due to its memory constraints and computational. Therefore, as illustrated in Figure \ref{fig:arquitetura}, the blockchain network is located above the process control level of the PAS hierarchy.

The blockchain network is strategically positioned from the supervision level as it is the last layer of IP-based communication required for interaction with the blockchain network. Thus, the levels of corporate management, plant management, supervision, and process control (just as a blockchain client) now interact in a decentralized manner. Process control level devices mediate communication between field devices and the top levels of PAS, through smart contracts that define task execution and raw data collection that can be used as assets for value creation to support business and manufacturing decision-making.

In this architecture, the blockchain network is used as middleware for communication between the levels of the PAS hierarchy. The blockchain network components can be defined as follows:
\begin{itemize}
    \item \textit{Blockchain Network}: set of nodes that perform a smart contract through a consensus algorithm.
    \item \textit{Node}: validates transactions and maintain consensus on the network, coordinating communication with other nodes (a node for the levels of process control, a node for plant management and corporate levels, and a node to the cloud). Each of the three nodes also has the following components:
    \begin{itemize}
    \item \textit{RestAPI}: allows programs of PAS hierarchy to interact with a validator using common HTTP/JSON standards.
    \item \textit{Smart Contract}: contains all factory operating logic in which it allows the input and search of data related to the various levels of the PAS hierarchy.
    \item \textit{Ledger}: each validator has its database in which it is used for data storage related to industrial operations.
    \end{itemize}
\end{itemize}

Smart contracts perform access security management for operator control and monitoring of PLC devices. Both PLC device management and control information is stored in the Ledger of each validator through transactions. Each transaction is identified by a unique address and can store states that are information represented in a compact data exchange format. States are used to represent information exchanged between levels of the PAS hierarchy. Two unique addresses are used to represent transactions and store states:
\begin{itemize}
    \item \textit{Device Log Address} (lo\_address): identifies all operations performed by the PLC, as well as the state changes of all field devices.
    \item \textit{Device Operations Address} (op\_address): identifies all operations to be performed by a PLC. Such operations are defined by the operator at the supervision level.
\end{itemize}

\subsection{Operation of Blockchain Network}

\begin{figure}[htp]
\centering
\includegraphics[width=\textwidth]{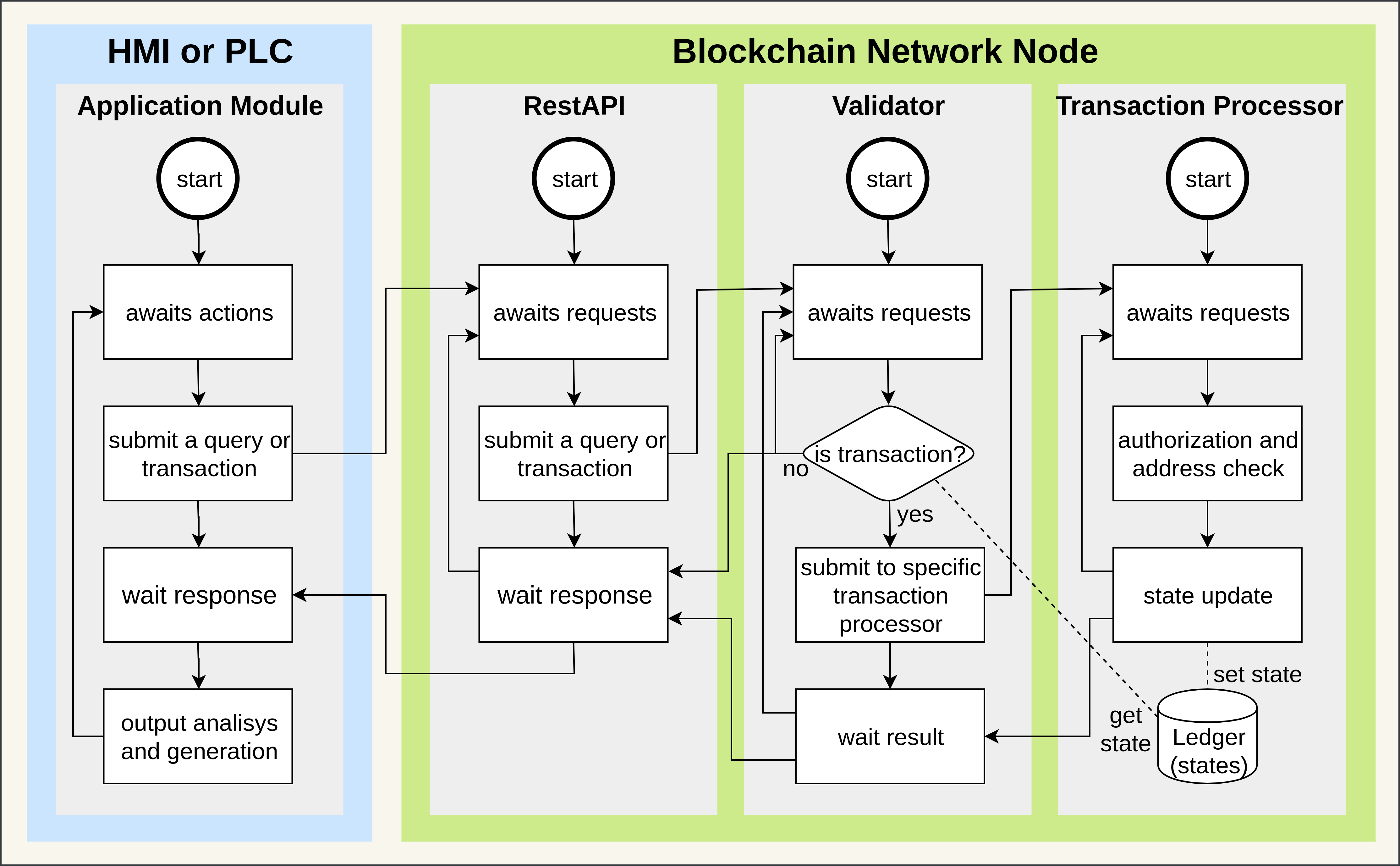}
\caption{Machine state diagram of the operation of each node of the blockchain network.}
\label{fig:m_e_b}
\end{figure}

As shown in Figure \ref{fig:m_e_b}, HMI (supervision level) or PLC (process control level) application modules can request queries or transactions to a blockchain network node through an HTTP request to RestAPI. Queries do not generate state changes in the ledger, so RestAPI receives this information directly from the validator. Already for transactions, changes are made in the ledger state, and thus, it is necessary to run the transaction processor for address analysis and storing state in the ledger.

The infrastructure of blockchain network nodes is deployed in strategic locations using edge computing concepts to reduce the latency in communication between each level to the given node \cite{garrocho2020blockchain}. For example, the shop floor is represented by the supervision node, where the HMI (supervision level) and PLC (process control level) devices communicate. The short and long term decision levels communicate with the management node. Despite this configuration, if a node fails, applications can communicate with another node, ensuring system operation.

\subsection{Process Control}

\begin{figure}[htp]
\centering
\includegraphics[width=\textwidth]{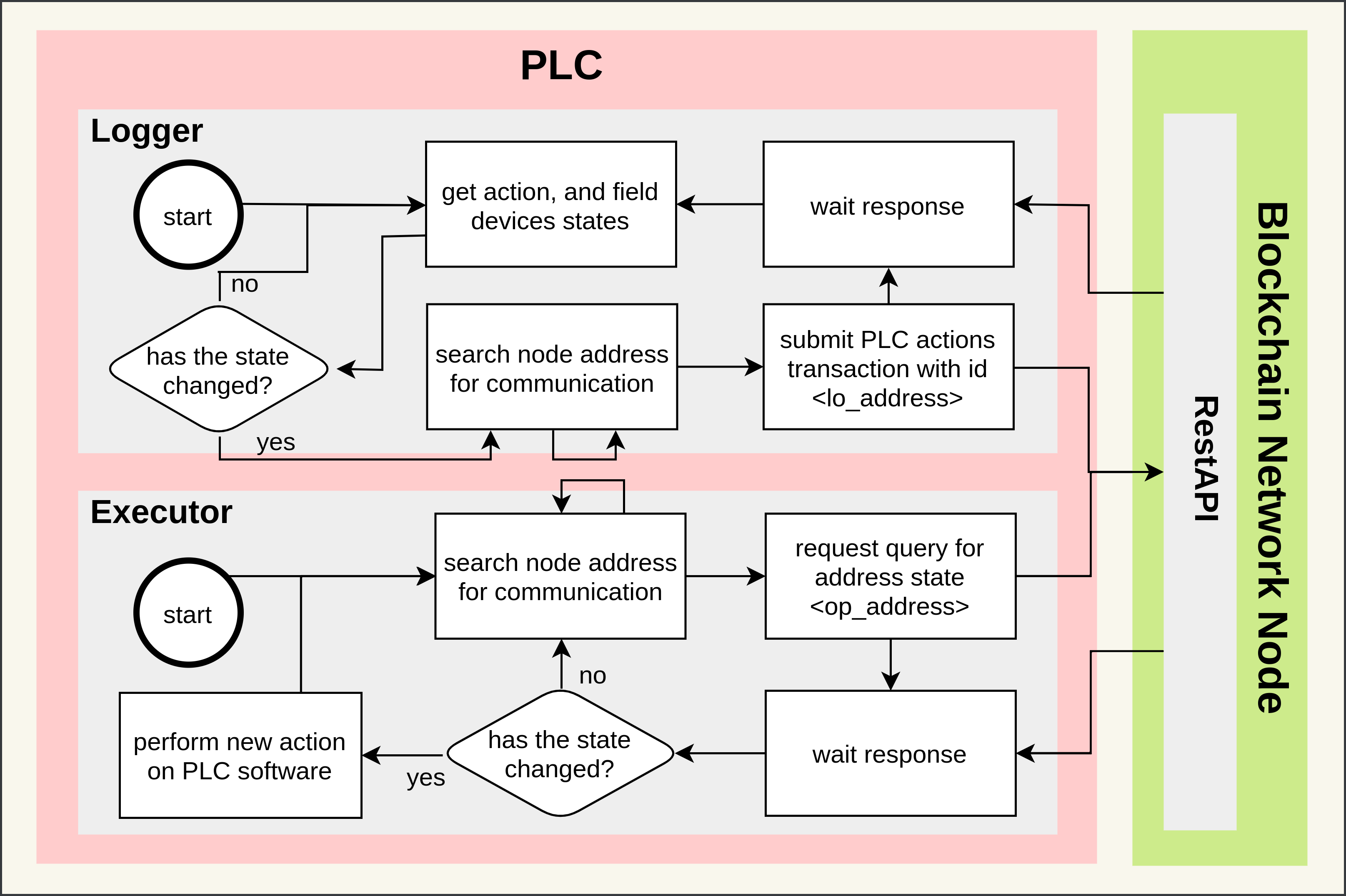}
\caption{State machine diagram of the operation of the PLC modules.}
\label{fig:m_e_p}
\end{figure}

Through PLC, it is possible to perform operations (e.g. compiling, executing and stopping programs) and monitoring field device states in real-time. Sensors and actuators are controlled and monitored by the process control level; however, these operations are defined by an operator at the supervision level. As illustrated in Figure \ref{fig:m_e_p}, two components represent the process control level:
\begin{itemize}
    \item \textit{Executor}: requests the current state (which is the operator submitted operation on the HMI) of the device operations address on the blockchain network. If the state is different from the last executed, a new action is performed by the PLC software.
    \item \textit{Logger}: collects the actions performed by the PLC and monitors through a ModbusTCP network the state changes of the PLC controlled field devices. After collection, and if there have been state changes, the data is submitted through the device log address on the blockchain network.
\end{itemize}

\subsection{Supervision}

\begin{figure}[htp]
\centering
\includegraphics[width=\textwidth]{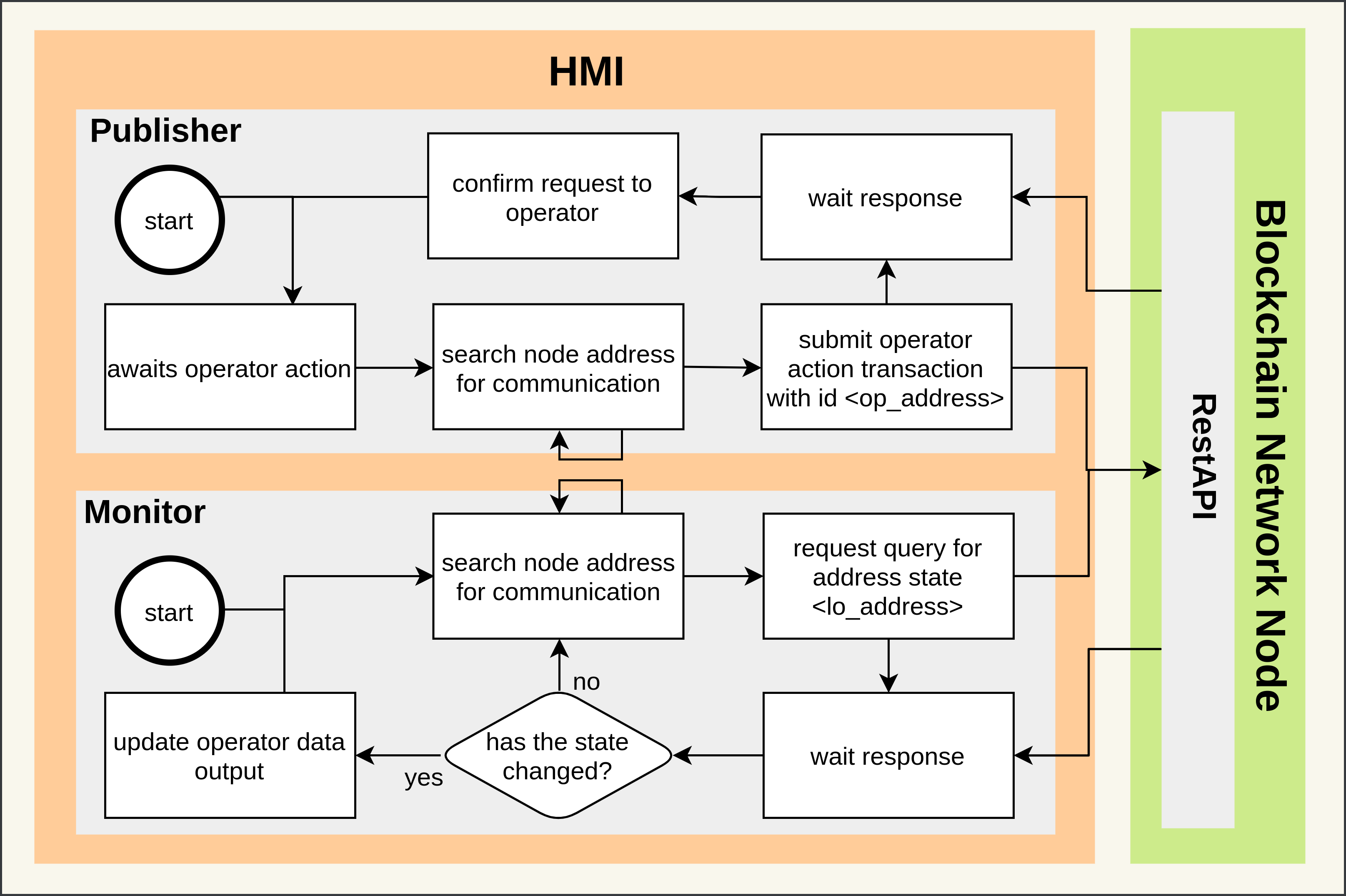}
\caption{State machine diagram of the operation of the HMI modules.}
\label{fig:m_e_s}
\end{figure}

At this level, HMI devices enable shop floor operators to track and interact with field device processes and information through the blockchain network. All operators performed by the operator are registered in the blockchain network and can be tracked. As illustrated in Figure \ref{fig:m_e_s}, two components represent the level of supervision:

\begin{itemize}
    \item \textit{Publisher}: publishes an operation (in data exchange format) chosen by the operator to the device operations address on the blockchain network. The operation contains the identification of the operator and the PLC device to perform the action.
    \item \textit{Monitor}: requests all states (which are the actions performed by PLC software and field devices) regarding the device log address on the blockchain network. If there are new states, the HMI output is updated to the operator.
\end{itemize}


\section{Proof of Concept and Evaluation}
\label{sec:ava}

To evaluate the proposed architecture, a motor control system was developed, which components are illustrated in Figure \ref{fig:prova_conceito}. The proposed architecture modules developed in this proof of concept do not affect PLC system performance; therefore, the rigorous requirements of real-time systems are not influenced. However, it is essential to evaluate the communication time between operator-generated tasks and performed by the PLC to assess the impact of delay on decision-making.

\subsection{Functioning of Proof of Concept and Technologies Used}

The Blockchain network consisted of a Linux server (Asus, 3.30 GHz Intel Core i7-5820K CPU, 32 GB RAM) running Hyperledger Sawtooth. At the supervisory level, a generic hardware was used because the focus was in desktop area, and  the modules were developed using the customer's sawtooth. At the process control level, the hardware used was a Raspberry Pi 3 board (4X ARM Cortex-A53 1.2 GHz, 1 GB RAM) as a PLC, while the software was OpenPLC\footnote{http://openplcproject.com} and the modules were developed using the Sawtooth client. At the field device level, we use a button and a Direct Current (DC) motor. 

For blockchain configuration, we adopted Hyperledger Sawtooth v1.1 for the following reasons: highly modular platform that separates the core system from the application level; supports permissionless and permissioned Infrastructure; allows parallel processing of transactions; Ethereum contract compatibility; pluggable consensus mechanisms; multilanguage support (Python, Javascript, Rust, C++, and Go); and byzantine-fault tolerant. For our proof of concept, we defined Sawtooth as permissioned infrastructure and Proof of Elapsed Time (PoET) is used as a consensus algorithm. In addition to the 3 Sawtooth nodes, we have also deployed a monitor node to collect real-time performance log data using InfluxDB\footnote{https://www.influxdata.com/}.

The 3-node blockchain network (representing PAS process control, supervision, and management levels) is used as middleware, in which the operator publishes commands and the PLC listens and executes commands in OpenPLC. PLC can, through OpenPLC, take a program (e.g. in Structured Text language) and compile it, or start and stop the execution of an already compiled program. Programs run by OpenPLC control the button and the DC motor. Any external action generated on the button will execute the DC motor. All state changes on field devices are monitored by the ModBusTCP network and recorded on the blockchain network by the PLC.

\begin{figure*}[h]
\centering
\includegraphics[width=\textwidth]{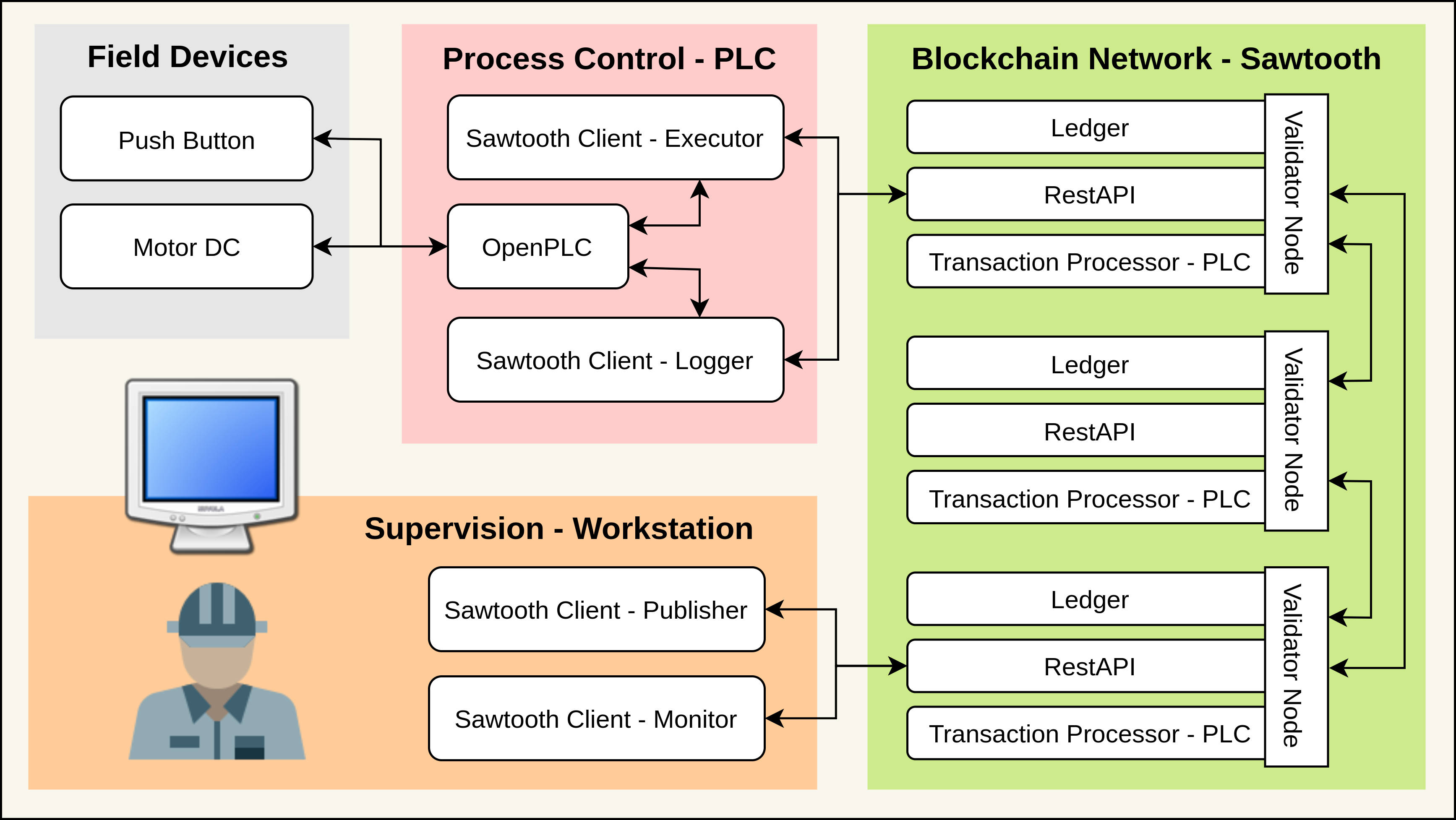}
\caption{Organization of the elements and technologies used in the proof of concept.}
\label{fig:prova_conceito}
\end{figure*}

Our implementation introduces a test environment that can be assembled using Docker Compose\footnote{https://docs.docker.com/compose/}, a tool for defining and running multi-container Docker applications. We build one container for each actor in the scenario shown in Figure \ref{fig:prova_conceito}. The Rest API, validator, and consensus engine run in separate containers and  assembled from images offered by Sawtooth. Already Sawtooth Clients (referring to architecture composed by HMI and PLC) and Transaction Processor were developed exclusively for these tests.

\subsection{Metrics and Measurement}

All experiments were conducted in an environment where the PLC device was directly connected to the Internet and the server was located 20 miles away. We measure 15 Mbps upload and download bandwidth between the PLC device and the blockchain network server using iperf\footnote{https://iperf.fr/} tool. All experiments was conducted to evaluate the impact of payload size (which is sensor and actuator data) on transaction response times. Payloads with sizes ranging from 0.5 kB to 750 kB were used. The maximum size chosen was the largest payload supported by the Hyperledger Sawtooth platform. The PLC device generated and submitted 2000 serial transactions to RestAPI from one of the three blockchain network nodes for each payload.

Based on the scenario presented, and following Hyperledger Performance and Scale Working Group guidelines \cite{hblock2018s}, we measured the following metrics:
\begin{itemize}
    \item \textit{Transaction Temporal Evolution}: total delay for creating, uploading, processing and committing a transaction:
    \begin{itemize}
        \item \textit{Transaction Creation}: total board delay for transaction preparation, hash generation, and payload coding.
        \item \textit{Transaction Upload}: total delay of the payload transfer from the card to RestAPI over the communication network.
        \item \textit{Transaction Latency}: total delay from transaction processing at RestAPI to confirmation that transaction has been committed by all nodes in the blockchain network.
    \end{itemize}
    \item \textit{Throughput}: also called transactions per second (tps), this metric represents the rate at which transactions are committed by all nodes in the blockchain network. This metric is defined in Equation 1.
\end{itemize}

\begin{equation}
Throughput = \dfrac{Total\hspace{0.1cm}Committed\hspace{0.1cm}Transactions}{Total\hspace{0.1cm}Time\hspace{0.1cm}Taken\hspace{0.1cm}in\hspace{0.1cm}Seconds}
\end{equation}

\section{Analysis and Discussion of Results}
\label{sec:ana}

\begin{figure*}[h]
\centering
\includegraphics[width=0.98\textwidth]{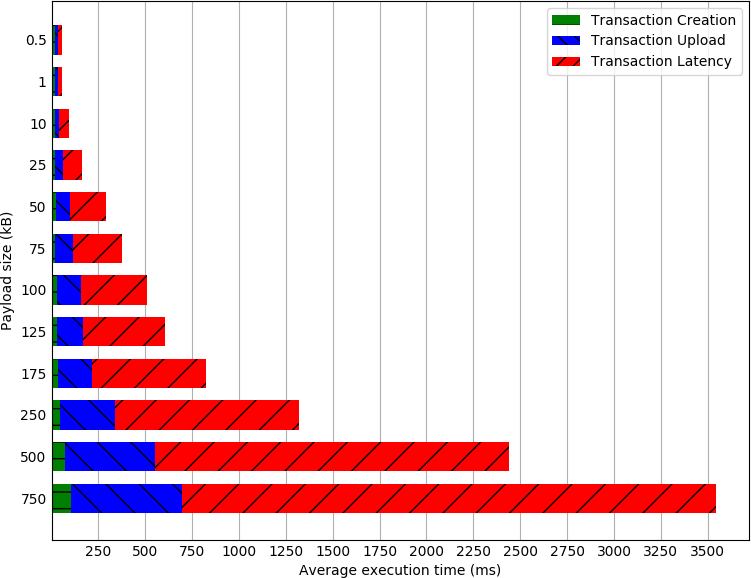}
\caption{Experimental results of transaction temporal evolution.}
\label{fig:result_temp}
\end{figure*}

Figure \ref{fig:result_temp} presents a graph showing the results of the transaction time evolution experiment, which is composed of the sub-operations (transaction creation, upload, and processing/commit) that represent the total time of the transaction. Figure \ref{fig:result_temp} shows a regressive curve in the total time required to complete a transaction, in which the payload size affects all three sub-operations.

The increase in time of transaction temporal evolution is related to the increase in the size of the payload. A larger payload implies a longer processing time to create the transaction on the board, sending data about the communication network and processing the transaction between the validating nodes. Considering the behaviour of the transaction latency sub-operation, payload size has a greater effect because operations are replicated between the 3 nodes of the blockchain network, generating a processing-related delay and consensus time between the validator nodes. 

\begin{figure*}[!h]
\centering
\includegraphics[width=0.985\linewidth]{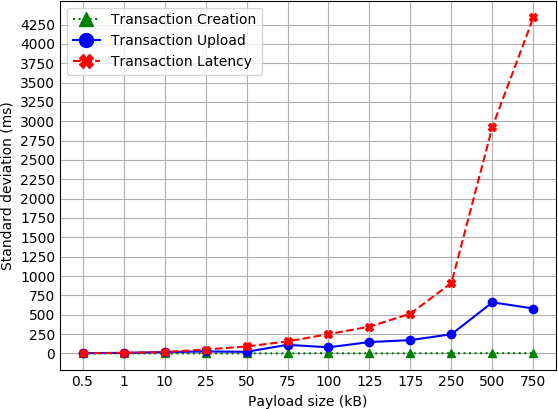}
\caption{Standard deviation of transaction temporal evolution.}
\label{fig:result_standard}
\end{figure*}

In addition to the results presented in Figure \ref{fig:result_temp}, Figure \ref{fig:result_standard} completely presents the results of the transaction temporal evolution with the standard deviation of sub-operations. The results show that payload size can also influence the standard deviation of times, generating a greater impact on the sub-operations of transaction upload and transaction latency.

The standard deviation for transaction creation is quite small because this variation is only related to interruptions caused by the board processor. As for transaction upload, the medium variation is related to collisions and packet losses in packet forwarding on an IP-based network. Finally, the high variation in transaction latency is related to consensus and replication of payloads between nodes in the blockchain network.

Thus, such results point to a high variability of the data, which makes the proposed approach unstable mainly for payloads greater than 100 kB. For payloads greater than 100 kB, each time an operation is performed, it will result in a high degree of unpredictability for the completion of the requested operation.

The results of the transaction temporal evolution experiment show a high standard deviation in which it can negatively influence industrial systems, especially real-time systems. Time variations in the transaction upload and transaction latency sub-operations can affect compliance with deadlines within which specific tasks must be completed. In industrial systems, this time is not suitable for processes where it can delay decision-making and compromise the system's time constraints.

In addition to the results presented in Figures \ref{fig:result_temp} and \ref{fig:result_standard} above, Table \ref{tabela:results} groups the means and standard deviation in a more broad and descriptive way. Compared to real-time M2M communication, the impact of transaction latency (shown in Table \ref{tabela:results}) is more significant, as real-time systems apply time requirements ranging from 10 to 100 ms. Thus, the results of these experiments showed that it is not possible to guarantee maximum times.

\begin{table}[!h]
\begin{tabular}{ccccccc}
\hline
\centering
\textbf{\begin{tabular}[c]{@{}c@{}}Payload\\Size (kB)\end{tabular}} & \multicolumn{2}{c}{\textbf{\begin{tabular}[c]{@{}c@{}}Transaction\\Creation (ms)\end{tabular}}} & \multicolumn{2}{c}{\textbf{\begin{tabular}[c]{@{}c@{}}Transaction\\Upload (ms)\end{tabular}}} & \multicolumn{2}{c}{\textbf{\textbf{\begin{tabular}[c]{@{}c@{}}Transaction\\Latency (ms)\end{tabular}}}} \\
\multicolumn{1}{c}{}                                            & \textbf{avg}          & \textbf{std dev}         & \textbf{avg}    & \textbf{std dev}    & \textbf{avg}     & \textbf{std dev}    \\ \hline
0.5                                                             & 16.88                      & 0.34                          & 16.32                & 2.65                     & 21.6                  & 4.97                     \\
1                                                               & 17                         & 0.36                          & 16.62                & 7.23                     & 24.15                 & 6.4                      \\
10                                                              & 18.19                      & 0.33                          & 21.86                & 17.18                    & 50.76                 & 19.47                    \\
25                                                              & 19.8                       & 0.35                          & 42.97                & 25.93                    & 102.11                & 49.78                    \\
50                                                              & 22.5                       & 0.89                          & 77.9                 & 20.95                    & 192.53                & 89.82                    \\
75                                                              & 19.57                      & 0.3                           & 93.38                & 111.26                   & 261.45                & 154.51                   \\
100                                                             & 28.36                      & 2.23                          & 103.01               & 79.04                    & 349.53                & 248.96                   \\
125                                                             & 30.9                       & 0.37                          & 139.21               & 146.79                   & 434.32                & 345.88                   \\
175                                                             & 36.72                      & 1.08                          & 180.48               & 172.06                   & 607.59                & 514.06                   \\
250                                                             & 45.47                      & 2.76                          & 339.76               & 247.62                   & 981.41                & 911.18                   \\
500                                                             & 73.56                      & 3.36                          & 513.63               & 660.34                   & 1888                  & 2930                     \\
750                                                             & 104.24                     & 4.32                          & 567.21               & 580.1                    & 2848                  & 4342                     \\ \hline
\end{tabular}
\caption{Classification and grouping of results of transaction temporal evolution.}
\label{tabela:results}
\end{table}

\begin{figure*}[!h]
\centering
\includegraphics[width=\linewidth]{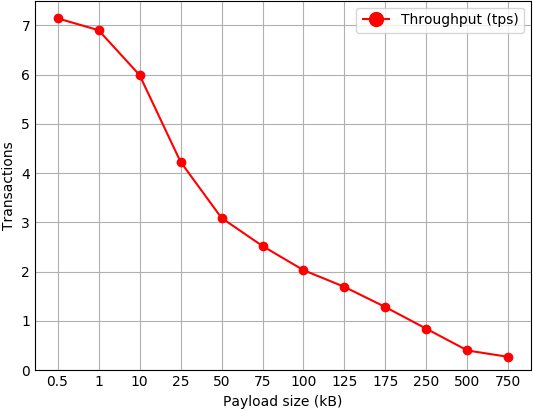}
\caption{Experimental results of throughput.}
\label{fig:result_throughput}
\end{figure*}

With higher payloads, a longer processing time of each validator's internal modules was verified.  Also with higher payloads, there is a longer delay in communication and message exchange between validator nodes for replication and transaction confirmation. Thus, in addition to impacting sub-transaction times on the time course of the transaction, payload sizes also affected throughput. Figure \ref{fig:result_throughput} illustrates the throughput results, which are mainly related to the results of the transaction latency sub-operation, which influences the number of transactions processed and committed per second in the blockchain network.

Therefore, the results of the experiments presented point out that one should not store large payloads that can negatively influence the total time variation for transaction processing. Thus, in our proposed architecture, one possible solution to this problem is just storing the payload hash of IIoT device data on the blockchain network, while the payload is stored outside the blockchain network (also called off-chain).

\section{Final Considerations}
\label{sec:con}

Current use and adoption of blockchain-based smart contract enforcement in Industry 4.0 are in its early stages, as this is an area that has much to explore. Cataloged solutions show that most related approaches are designed for specific processes that are intended to automate horizontal communication of PAS. In this new context, we designed and developed a vertical integration approach for PAS based on blockchain and smart contracts.

The introduction of the proposed approach for PAS vertical integration has resulted in full process decentralization and automated communication across the supply chain. Also, test results have shown in real experiments that many delays primarily related to transaction submission and processing have a high and variable time that is sometimes unsuitable for real-time M2M communications, requiring alternatives (e.g. off-chain data storage) to meet system requirements in real-time.

For future work, new assessments will be extended in environments with a more significant number of PLC devices to assess the scalability and behavior of the proposed architecture. Besides, we will use simulators to perform such assessments, which will provide an assessment scenario in which to allow understanding of various aspects of the proposed architecture operation in the industrial environment.

\section*{Acknowledgment}
We acknowledge the support of the Brazilian research agencies National Council for Scientific and Technological Development (CNPq) and Coordination for the Improvement of Higher Education Personnel (Capes), the Minas Gerais State Research Foundation (FAPEMIG), the Minas Gerais Federal Institute of Education, Science and Technology (IFMG), and the Federal University of Ouro Preto (UFOP).


\bibliography{mybibfile}

\end{document}